\documentclass[twocolumn,superscriptaddress,aps,preprintnumbers,amsmath,amssymb,prl,nofootinbib]{revtex4}

\usepackage{graphicx}
\usepackage{epstopdf}
\usepackage{dcolumn}
\usepackage{bm}
\usepackage{hyperref}
\usepackage{color}

\begin{document}


\def\a{\alpha}
\def\b{\beta}
\def\c{\varepsilon}
\def\d{\delta}
\def\e{\epsilon}
\def\f{\phi}
\def\g{\gamma}
\def\h{\theta}
\def\k{\kappa}
\def\l{\lambda}
\def\m{\mu}
\def\n{\nu}
\def\p{\psi}
\def\q{\partial}
\def\r{\rho}
\def\s{\sigma}
\def\t{\tau}
\def\u{\upsilon}
\def\v{\varphi}
\def\w{\omega}
\def\x{\xi}
\def\y{\eta}
\def\z{\zeta}
\def\D{\Delta}
\def\G{\Gamma}
\def\H{\Theta}
\def\L{\Lambda}
\def\F{\Phi}
\def\P{\Psi}
\def\S{\Sigma}

\def\o{\over}
\def\beq{\begin{eqnarray}}
\def\eeq{\end{eqnarray}}
\newcommand{\gsim}{ \mathop{}_{\textstyle \sim}^{\textstyle >} }
\newcommand{\lsim}{ \mathop{}_{\textstyle \sim}^{\textstyle <} }
\newcommand{\vev}[1]{ \left\langle {#1} \right\rangle }
\newcommand{\bra}[1]{ \langle {#1} | }
\newcommand{\ket}[1]{ | {#1} \rangle }
\newcommand{\EV}{ {\rm eV} }
\newcommand{\KEV}{ {\rm keV} }
\newcommand{\MEV}{ {\rm MeV} }
\newcommand{\GEV}{ {\rm GeV} }
\newcommand{\TEV}{ {\rm TeV} }
\newcommand{\1}{\mbox{1}\hspace{-0.25em}\mbox{l}}
\def\diag{\mathop{\rm diag}\nolimits}
\def\Spin{\mathop{\rm Spin}}
\def\SO{\mathop{\rm SO}}
\def\O{\mathop{\rm O}}
\def\SU{\mathop{\rm SU}}
\def\U{\mathop{\rm U}}
\def\Sp{\mathop{\rm Sp}}
\def\SL{\mathop{\rm SL}}
\def\tr{\mathop{\rm tr}}

\def\IJMP{Int.~J.~Mod.~Phys. }
\def\MPL{Mod.~Phys.~Lett. }
\def\NP{Nucl.~Phys. }
\def\PL{Phys.~Lett. }
\def\PR{Phys.~Rev. }
\def\PRL{Phys.~Rev.~Lett. }
\def\PTP{Prog.~Theor.~Phys. }
\def\ZP{Z.~Phys. }

\def\dd{\mathrm{d}}
\def\ff{\mathrm{f}}
\def\BH{{\rm BH}}
\def\inf{{\rm inf}}
\def\ev{{\rm evap}}
\def\eq{{\rm eq}}
\def\SM{{\rm sm}}
\def\Mpl{M_{\rm Pl}}
\def\GeV{{\rm GeV}}
\newcommand{\Red}[1]{\textcolor{red}{#1}}

\title{
Large scale cosmic perturbation from 
evaporation of primordial black holes
}

\author{Tomohiro Fujita}
\affiliation{Kavli IPMU (WPI), TODIAS, University of Tokyo, Kashiwa, 277-8583, Japan}
\author{Keisuke Harigaya}
\affiliation{Kavli IPMU (WPI), TODIAS, University of Tokyo, Kashiwa, 277-8583, Japan}
\author{Masahiro Kawasaki}
\affiliation{ Institute for Cosmic Ray Research, University of Tokyo, Kashiwa 277-8582, Japan}
\affiliation{Kavli IPMU (WPI), TODIAS, University of Tokyo, Kashiwa, 277-8583, Japan}

\begin{abstract}
We present a novel mechanism to generate the 
cosmic perturbation from  evaporation of primordial black holes.
A mass of a field is fluctuated if it is given by a vacuum expectation value of a light scalar field because of the quantum fluctuation during inflation. The fluctuated mass causes variations of the evaporation time
of the primordial black holes.
Therefore provided the primordial black holes dominate the Universe when they evaporate,
primordial cosmic perturbations are generated.
We find that the amplitude of the large scale curvature perturbation generated 
in this scenario can be consistent with the  observed value.
Interestingly, our mechanism works even if all fields
that are responsible for inflation and the generation of the cosmic perturbation
are decoupled from the visible sector
except for the gravitational interaction.
An implication to the running spectral index is also discussed.
\end{abstract}
\maketitle
\preprint{IPMU 13-0132}
\preprint{ICRR-report-656-2013-5}

%
\section{I.  Introduction}
%

Recent observations of the cosmic microwave background
radiation determine the cosmological parameters with increased accuracy
and hence we know the amplitude and the tilt of the initial cosmological
perturbations on the large scale~\cite{Ade:2013zuv}.
However, the mechanism which generates
such perturbations is still unknown. 
Even if one assumes that inflation occurs and it stretches the quantum fluctuations of light scalar fields into the cosmological ones~\cite{Guth:1980zm}, what actually produces the observed perturbations is quite obscure.
For example, in curvaton~\cite{Mollerach:1989hu} or modulated reheating~\cite{Dvali:2003em} scenarios, the scalar field which is responsible for the large scale cosmic perturbations is different from the inflaton itself.
In addition, all known scenarios, such as single field inflation, curvaton and modulated reheating, are not necessarily the only possibilities to be considered.
Thus it is important to investigate another feasible mechanism generating the perturbation, in order to understand what actually happened in the early Universe.

Primordial black holes (PBHs) are black holes which formed in the early Universe.
After the pioneer work by Zel'dovich and Novikov~\cite{Zeldovich:1967}
in 1966, PBHs have attracted much attention for a long time.
Now it is known that PBHs can be produced by
large density perturbations due to inflation
or preheating~\cite{Taruya:1998cz} as well as a sudden reduction in the pressure~\cite{Jedamzik:1996mr},
bubble collisions~\cite{Crawford:1982yz}, collapses of cosmic strings~\cite{Hogan:1984zb}, and so on~\cite{Khlopov:2008qy}. PBHs with small masses, 
$M_{\rm BH} \lesssim 10^{15}$g, evaporate until today
by the  Hawking radiation~\cite{Hawking:1974sw} and
lead to rich phenomenologies including
entropy productions, baryogenesis~\cite{Turner:1979bt} and neutrino radiations~\cite{Bugaev:2000bz}.
On the other hand, PBHs with large masses, $M_{\rm BH} \gtrsim 10^{15}$g,
survive today and they can be the candidate of cold dark matter and
might affect the large scale structure~\cite{Meszaros:1975ef},
possibly seeding the supermassive black holes~\cite{Carr:1984id}.
However, although a number of earlier works investigate
the possible roles of PBHs, 
PBHs have never been studied as
the origin of the cosmic perturbation.

Before showing detailed calculations, let us briefly explain
the basic idea of our mechanism where the fluctuation
of the PBH evaporation time generates cosmic perturbations:
First of all, 
we 
assume that one field ($\psi$)
acquires its mass $m_\psi$ from another field (light scalar field 
$\phi$) 
by the Higgs mechanism.
Then the mass of $\psi$, $m_\psi$,
is fluctuated if the field value of the light scalar field 
$\phi$ is also fluctuated by inflation.
Second, a lifetime of a PBH depends on 
$m_\psi$
if $m_\psi$ 
is larger than the initial Hawking temperature
$T_{\rm BH}$ of the PBH.
It can be understood step by step.
Since the Hawking temperature 
$T_{\rm BH}$ 
is inversely proportional to the PBH mass
$M_{\rm BH}$, $T_{\rm BH}$ increases as $M_{\rm BH}$ decreases due to the Hawking radiation.
Next, the PBH can emit particles whose masses are 
smaller than $T_{\rm BH}$.
Thus the number of particle species in the Hawking radiation
rises as $T_{\rm BH}$ increases. 
Moreover the energy loss rate of the PBH is proportional to
the degree of freedom of the radiated particles.
Therefore the energy loss of the PBH
accelerates when  
$T_{\rm BH}$ exceeds $m_\psi$.
As a result, the variation of $m_\psi$
causes the variation of the time of the PBH evaporation.
Finally, if PBHs dominate the Universe, the fluctuation of the PBH evaporation time is nothing but the fluctuation of the (second) reheating time.
\footnote{Here we consider that the first reheating occurs after
inflation and refer to the PBH evaporation as the second reheating because
the radiation component from the inflaton become negligible after PBHs dominate the Universe. }
Thus cosmic perturbations are generated via the PBH evaporation.

This novel mechanism has several interesting features.
First, the generation of perturbations by this mechanism
is general because it is natural to expect that an unknown particle acquires its mass by the Higgs mechanism, if some symmetry
forbids the mass term.
Second, this mechanism is still viable even if all fields which are relevant to 
inflation and the generation of the cosmic perturbations
are decoupled from the visible sector 
except for the gravitational interaction.
It is because particles in the visible sector are emitted by the Hawking radiation
even if PBHs are formed from an invisible sector.
Third, the PBH dominated universe leads to 
the rich phenomenologies as we have mentioned.
It would be interesting to explore these scenarios
in combination with our mechanism.
In this article, however, 
we focus on the generation mechanism 
of cosmic perturbations via PBHs.

This article is organized as follows.
First, we explain the generation mechanism in detail and
show that the observed magnitude of the cosmic perturbations can be
realized.
Next, we make a prediction of the running spectral index
when the PBHs are produced due to a blue-tilted perturbation generated
by the inflaton.
We show that the negatively large running is predicted.
The final section is devoted to conclusions and discussion.

%
\section{II.  Perturbation of PBH lifetime}
%

In this section, we compute the curvature perturbation
produced by the PBH evaporation when a mass of a
field is fluctuated by the Higgs mechanism.
A PBH loses its mass due to the Hawking radiation~\cite{Hawking:1974sw} 
which is characterized by the Hawking temperature,
\begin{equation}
T_\BH = \frac{\Mpl^2}{M_\BH},
\end{equation}
where $M_\BH$ is the mass of the PBH and 
$\Mpl \approx 2.43\times10^{18}$GeV is the reduced Planck mass.
The mass loss rate is given by
\begin{equation}
\frac{\dd M_\BH}{\dd t}
=
-\frac{\pi^2}{120}g_* T_\BH^4 \times 4\pi r_s^2
=
-\frac{\pi}{480} g_* \frac{\Mpl^4}{M_\BH^2},
\label{EoM of PBH mass}
\end{equation}
where $g_*$ is the effective degrees of freedom of the radiated particles and
$r_s \equiv M_\BH/4\pi \Mpl^2$ is the Schwarzschild radius.
\footnote{A black hole not only emits particles 
but also absorbs the radiated particles 
by its gravitational attraction. 
This effect can be taken into account as overall factors (grey body factors)
of the momentum distribution functions of the radiated particles
and then eq.~(\ref{EoM of PBH mass}) is modified.
Nevertheless we ignore this effect for simplicity,
since it alters the following estimations only by few factors.}
Let us consider a case where a field $\psi$
is added to the standard model. We define $m_\psi$ as the mass of $\psi$
and $g_\psi$ as its effective degrees of freedom.
Then $g_*$  varies approximately as
\begin{equation}
g_*
= \left\{ 
\begin{array}{ll}
g_\SM & ( T_\BH < m_\psi ) \\
g_\SM + g_\psi & ( T_\BH > m_\psi)
\end{array}\right.,
\label{g change}
\end{equation}
where $g_\SM= 106.75$ is the total degrees of freedom in the standard model.
One can find the solution of Eq.~(\ref{EoM of PBH mass})
under Eq.~(\ref{g change}) as
\begin{equation}
\tau =
\tau_\SM
\left[
1 -
\left(\frac{T_0}{m_\psi}\right)^3
\frac{g_\psi}{g_\SM + g_\psi}
\right],
\ 
\tau_\SM \equiv \frac{160\, M_0^3}{\pi\, g_\SM \Mpl^4},
\label{PBH life time}
\end{equation}
where $m_\psi$ is assumed to be larger than the initial
Hawking temperature $T_0$.
Otherwise, $\psi$ is emitted from the beginning and then
$\tau$ does not depend on $m_\psi$.
On the other hand, 
$\psi$ makes a small difference if $m_\psi\gg T_0$
because a black hole evaporates rapidly at $t\simeq \tau$.

Next, we introduce a scalar field $\phi$ and assume that
$\psi$ is a fermion field.
We suppose that the interaction between $\phi$ and $\psi$
is given by the Yukawa coupling
and it gives $\psi$ the mass $m_\psi$, 
\begin{equation}
\mathcal{L}_{\rm int} =- y \phi \bar{\psi} \psi
\quad
\Longrightarrow
\quad 
m_\psi \equiv y\phi\,,
\label{psi mass}
\end{equation}
where $y$ is a Yukawa coupling constant.
Provided that $\psi$ does not acquire any significant mass except for Eq.~(\ref{psi mass}),
the fluctuation of $\phi$ which is generated during inflation 
perturbs the mass of $\psi$.
Then it causes the perturbation of the PBH lifetime $\tau$ 
through Eq.~(\ref{PBH life time}). One can find that the perturbation of $\tau$,
\begin{equation}
\delta\tau = 
\tau_\SM 
\frac{3g_\psi}{g_\SM + g_\psi}
\left(\frac{T_0}{m_\psi}\right)^3
\frac{\delta\phi}{\phi},
\end{equation}
where $\delta\phi$ is the fluctuation of $\phi$
and the contribution of $\phi$ to $g_*$ is ignored.
Note that in order for $\delta \phi$ 
to survive until the PBH evaporation,
$\phi$ should not gain a large thermal mass 
by interactions with the radiation
which is originated from the inflaton.

Let us evaluate the resultant curvature perturbation.
Here we assume that the PBHs dominate the Universe before 
their evaporation.
The Universe is in the matter dominated era before the PBH
evaporation and
enters 
the radiation dominated era after that.
Then the curvature perturbation $\zeta$
generated at the evaporation of the PBHs
can be calculated in the same way as the modulated reheating cases in which
$\zeta$ is given by $\zeta_{\rm MR}=-\delta\Gamma/(6\Gamma)$,
where $\Gamma$ is the decay rate of the inflaton~\cite{Dvali:2003em}.
We derive this formula in the Appendix.
Since in the PBH evaporation case $\Gamma$ corresponds to 
the inverse of the PBH lifetime $\tau^{-1}$,
$\zeta$ generated by the PBH lifetime perturbation is given by
\begin{equation}
\label{eq:zeta}
\zeta =
\frac{\delta\tau}{6\tau}
=
\frac{1}{2}\frac{g_\psi}{g_\SM + g_\psi}
\left(\frac{T_0}{m_\psi}\right)^3
\frac{\delta\phi}{\phi},
\end{equation}
where $\tau_\SM \gg \delta\tau$ is assumed.
Provided that $\phi$ is light during inflation and the power spectrum of its fluctuation is $\mathcal{P}_{\delta\phi}= H_\inf/2\pi$,
the power spectrum of the curvature perturbation is given by
\begin{equation}
\mathcal{P}_\zeta^{1/2}
=
\frac{y}{4\pi}\frac{g_\psi}{g_\SM + g_\psi}
\frac{T_0^3 H_\inf}{m_\psi^4}
\,,
\label{power spectrum 1}
\end{equation}
where $H_\inf$ is the Hubble parameter during inflation.

Although several generation mechanisms of PBHs are proposed,
we simply assume that the PBHs form right after the end of inflation
without specifying a concrete model.
In that case, the typical mass of the PBHs is evaluated as
\begin{equation}
M_0 = \gamma M_{\rm horizon}^3
= 4\pi\gamma \frac{\Mpl^2}{H_\inf},
\label{initial mass}
\end{equation}
where $M_{\rm horizon}$ denotes the energy density in the horizon volume at the end of inflation and $\gamma \approx 0.2$ is the numerical factor representing
the effect that the pressure of radiations prevents the structure formation~\cite{Carr:1975qj}.
The initial Hawking temperature is
\begin{equation}
T_0 = \frac{H_\inf}{4\pi\gamma}
\approx 0.4 H_\inf \,.
\label{GH temperature}
\end{equation}
%
Substituting Eq.~(\ref{GH temperature}) into Eq.~(\ref{power spectrum 1}),
we obtain
\begin{equation}
\mathcal{P}_\zeta^{1/2}
=
\frac{y}{(4\pi)^4 \gamma^3}\frac{g_\psi}{g_\SM + g_\psi}
\left(\frac{H_\inf}{m_\psi}\right)^4 .
\label{power spectrum 2}
\end{equation}
When $g_\psi \ll g_\SM$, Eq.~(\ref{power spectrum 2}) is evaluated as
\begin{equation}
\mathcal{P}_\zeta^{1/2}
\approx
10^{-5}\times
\left(\frac{y}{0.3}\right)
\left(\frac{g_\psi}{1}\right)
\left(\frac{H_\inf}{m_\psi}\right)^4
\,.
\end{equation}

Let us discuss whether the observed curvature perturbation, ${\cal
P}_\zeta^{1/2}\sim 10^{-5}$,
can be realized by the evaporation of the PBHs.
To obtain the observed curvature perturbation, $m_{\psi}$ must
be close to $H_{\rm inf}$. Otherwise, the Yukawa
coupling $y$ must be larger than the unitarity bound $\sim4\pi$.
However, the required coincidence of $m_{\psi}$ and $H_{\rm inf}$
is not a fine-tuned one; it is just within a factor of few.

The coincidence can be realized more naturally in the following way.
If $\phi$ couples to several fields, as is the case for the standard model Higgs, it is not unnatural that one of these fields has a mass close to the Hubble scale within a factor of a few.
Note also that the perturbation depends on the degree of freedom of
the fermions, $g_\psi$.
If $\phi$ couples to several fermions with the same Yukawa coupling,
which can be guaranteed by assuming a flavor symmetry among the fermions, the required closeness of the mass scales is relaxed.

Before closing this section, let us note the condition
in which the PBHs dominate the Universe prior to their evaporation.
Provided that the PBHs form right after the end of inflation and
the inflaton oscillation phase is negligible (instant reheating),
the density parameter of the PBHs at their formation epoch,
$\beta\equiv \Omega_{\rm PBH}(t_{\rm form})$,
is constrained as
\begin{equation}
\beta 
\ >\ 
\sqrt{\frac{g_\SM}{20480 \pi^2 \gamma^3}}\frac{H_\inf}{\Mpl}
\ \approx\
10^{-8}
\left(\frac{H_\inf}{10^{11}\GeV}\right),
\label{PBH dominant condition}
\end{equation}
where Eq.~(\ref{initial mass}) is used.
If $\beta$ is smaller than the lower bound in Eq.~(\ref{PBH dominant condition}), 
the resultant curvature power spectrum $\mathcal{P}_\zeta$
decreases by a factor
of $\left( 3\Omega_{\rm PBH}(\tau)/(4-\Omega_{\rm PBH}(\tau)) \right)^2$.

\section{IV. Implication to the running spectral index}
%

In this section, we briefly discuss the prediction of the running
spectral index when the PBHs are produced due to a blue-tilted
perturbation from the inflaton. To be concrete, we discuss this based on the
hybrid inflation~\cite{Linde:1991km}, in which a blue-tilted spectral
is easily obtained.%
\footnote{The following analysis is essentially true for other inflaton potentials, if  only $\eta$ is significantly large among the slow-roll parameters.}
We assume the following standard hybrid inflaton potential:%
\begin{eqnarray}
 V(s) = V_0 + \frac{1}{2}m_s^2 s^2+\cdots,
\end{eqnarray}
where $s$ is the inflaton and $\cdots$ includes the interactions with
the waterfall sector.

With a simple calculation, we obtain the relation%
\begin{eqnarray}
\eta = \frac{1}{2N_*} {\rm ln}\left({\cal P}_{\zeta e}^{\rm inf}/{\cal P}_{\zeta*}^{\rm inf}\right),
\end{eqnarray}
where ${\cal P}_\zeta^{\rm inf}$, $\eta$ and $N$ are the curvature
perturbation generated by the inflaton,
the second slow-roll parameter and the number of e-foldings, respectively. The indices
$*$ and $e$ denote that the value is evaluated at the horizon exit of
the scale of the interest and
the end of the inflation, respectively.
Note that the curvature perturbation should be large enough at the small
scale in order to produce the PBHs, and should be small at the large scale,
\begin{eqnarray}
 {\cal P}_{\zeta e}^{\rm inf} = {\cal O}(1),\qquad
 {\cal P}_{\zeta *}^{\rm inf} < 10^{-10}.
\end{eqnarray}
Therefore we obtain the lower bound for $\eta$,
\begin{eqnarray}
 \eta > 0.2 \frac{60}{N_*}.
\end{eqnarray}
Such large $\eta$ is natural
in the supergravity theory~\cite{Ovrut:1983my}.

On the other hand, the spectral index $n_s$ of the perturbation
generated by the PBH evaporation is
given by $n_s = 1- 2\epsilon_*$,
since the perturbation is originated from a light field $\phi$
other than the inflaton.
If the mass of $\phi$ is so large that it affects the
spectral index, $\phi$ would begin an oscillation before the evaporation
of the PBHs.
To obtain the value consistent with the Planck results~\cite{Ade:2013zuv}, $n_s
= 0.9607 \pm 0.0063$~(95\% C.L.), $\epsilon_* = 0.020 \pm 0.003$ is required.

In the end, we obtain the prediction of the running of the spectral
index,
\begin{eqnarray}
 \frac{\dd n_s}{\dd \ln k} \simeq -4 \epsilon_* \eta + 8\epsilon_*^2 < -0.011\frac{60}{N_*}.
\end{eqnarray}
Thus,  a large running is easily obtained. 

%
\section{IV. Conclusions and discussion}
%
In this article, we have proposed a new generation mechanism of cosmic perturbations from the evaporation of PBHs.
It has been shown that the mechanism is compatible with the observed
magnitude of the curvature perturbation.
The implication to the running spectral index has also been discussed.

As has been mentioned in the Introduction, the generation mechanism of cosmic
perturbations from the evaporation of PBHs has an interesting feature.
Even if all fields responsible for inflation and cosmic
perturbations are decoupled from the visible sector, this mechanism is viable.

%
\section{Acknowledgements}
%
This work is supported by Grant-in-Aid for Scientific Research 
from the Ministry of Education, Science, Sports, and Culture
(MEXT), Japan, No. 25400248 (M.K.), No. 21111006 (M.K.)
and also by World Premier International Research Center Initiative
(WPI Initiative), MEXT, Japan.
T.F. and K.H. acknowledge the support by JSPS Research
Fellowship for Young Scientists.

\appendix

\section{APPENDIX: CURVATURE PERTURBATION AND THE EVAPORATION RATE}
In this appendix, we derive the formula for the curvature perturbation
when the evaporation rate of the PBHs, $\Gamma$, fluctuates, with the aid of the so-called $\delta N$
formula~\cite{Sasaki:1995aw}.
We assume that the PBHs dominate the Universe before their evaporation.

We take a flat time slice $t_i$ well before the PBHs evaporate but well after the
PBHs dominate the Universe as the initial time slice.
We also take a uniform density time slice $t_f$ well after the PBHs
evaporate as the final time slice.
Note that the Universe is in the matter dominated era before the PBH evaporation and enters the radiation dominated era after that.
Therefore, the number of e-foldings between the two slices is
given by
\begin{eqnarray}
\label{eq:efold}
 N = {\rm
  ln}\left[\left(\frac{\tau}{t_i}\right)^{2/3}\left(\frac{t_f}{\tau}\right)^{1/2}\right]
  = \frac{1}{6}{\rm ln}\tau + {\rm const},
\end{eqnarray}
where $\tau$ is the lifetime of the PBHs, $\tau=\Gamma^{-1}$.
By taking a variation of Eq.(\ref{eq:efold}), we obtain
\begin{eqnarray}
 \zeta = \delta N = \frac{1}{6}\frac{\delta\tau}{\tau}=-\frac{1}{6}\frac{\delta\Gamma}{\Gamma},
\end{eqnarray}
which is used in Eq.(\ref{eq:zeta})

\end{document}